\documentclass[journal=jacsat,manuscript=article]{achemso}
\usepackage{chemformula}
\usepackage[T1]{fontenc}

\author{Josh Vogwell$^{1}$}
\author{Laura Rego$^{1,2}$}
\author{Olga Smirnova$^{3,4}$}
\author{David Ayuso$^{1,3}$}
\email{d.ayuso@imperial.ac.uk}
\affiliation{$^1$Department of Physics, Imperial College London, SW7 2AZ London, United Kingdom \\
$^2$University of Salamanca, 37008 Salamanca, Spain \\
$^3$Max-Born-Institut, Max-Born-Str. 2A, 12489 Berlin, Germany \\
$^4$Technische Universit\"at Berlin, 10623 Berlin, Germany}

\title{Ultrafast control over chiral sum-frequency generation}

\begin{document}

\begin{abstract}

We introduce an ultrafast all-optical approach for efficient chiral recognition which relies on the interference between two low-order nonlinear processes which are ubiquitous in nonlinear optics: sum-frequency generation and third-harmonic generation.
In contrast to traditional sum-frequency generation, our approach encodes the medium's handedness in the \emph{intensity} of the emitted harmonic signal, rather than in its \emph{phase}, and it enables full control over the enantiosensitive response.
We show how, by sculpting the sub-optical-cycle oscillations of the driving laser field, we can force one molecular enantiomer to emit bright light while its mirror twin remains dark, thus reaching the ultimate efficiency limit of chiral sensitivity via low-order nonlinear light-matter interactions.
Our work paves the way for ultrafast and highly efficient imaging and control of the chiral electronic clouds of chiral molecules using lasers with moderate intensities, in all states of matter: from gases to liquids to solids, with molecular specificity and on ultrafast timescales.

\end{abstract}

\section{Introduction}

Chirality is a universal type of asymmetry that naturally arises in molecules, optical fields, viruses, or even galaxies.
Just like a chiral glove would either fit our left or right hand, but not both, the two non-superimposable mirror-reflected versions of a chiral molecule (enantiomers) can behave very differently when they interact with another chiral entity, e.g. another chiral molecule.
Since the majority of molecules supporting biological life are chiral, methods for detecting, quantifying and manipulating molecular chirality are of great importance and interest, particularly in biochemical and pharmaceutical contexts.

Traditional chiroptical methods \cite{Berova2007,Nafie1976,Schellman1975,Polavarapu1998} 
rely on weak linear effects which arise beyond the electric-dipole approximation, posing significant limitations for detecting dilute samples and ultrafast spectroscopy \cite{Oppermann2019}.
This challenge\cite{Ayuso2022PCCP_persp} can be addressed by developing and applying approaches which rely exclusively on the molecular response to the local polarisation of the electric-field vector of an electromagnetic wave\cite{Ayuso2022PCCP_persp,Ordonez2018PRA,Ritchie1976PRA,Powis2000JCP,Bowering2001PRL,Garcia2003JCP,Garcia2013NatComm,Janssen2014PCCP,Lux2012Angew,Lehmann2013JCP,Lux2015ChemPhysChem,Kastner2016CPC,Comby2016JPCL,Beaulieu2016FD,Comby2018NatComm,Beaulieu2018NatPhys,Demekhin2018PRL,Goetz2019PRL,Rozen2019PRX,Ordonez2022PCCP,Yachmenev2019PRL,Milner2019PRL,Patterson2013,Eibenberger2017PRL,Shubert2016,Perez2017,Eibenberger_PRL_2022,Fischer2000PRL,Belkin2000PRL,Belkin2001PRL,Fischer2002CPL,Fischer2003PRL,Fischer2005,Okuno2018JCP,Neufeld2019PRX,Ayuso2019NatPhot,Ayuso2021NatComm,Ayuso2021Optica,Ayuso2022PCCP,Ayuso2022OptExp,Rego2023,Khokhlova2022}.
One strategy is to record the forward-backward asymmetry in the photo-electron angular distributions upon ionisation with circularly polarised light\cite{Ritchie1976PRA,Powis2000JCP,Bowering2001PRL,Garcia2003JCP,Garcia2013NatComm,Janssen2014PCCP,Lux2012Angew,Lehmann2013JCP,Lux2015ChemPhysChem,Kastner2016CPC,Comby2016JPCL,Beaulieu2016FD,Comby2018NatComm,Beaulieu2018NatPhys}. 
This method, originally proposed \cite{Ritchie1976PRA} and implemented 
\cite{Powis2000JCP,Bowering2001PRL,Garcia2003JCP,Garcia2013NatComm,Janssen2014PCCP} in the \emph{linear} regime, has been extended to the \emph{nonlinear}\cite{Janssen2014PCCP,Lux2012Angew,Lehmann2013JCP,Lux2015ChemPhysChem,Kastner2016CPC,Comby2016JPCL,Beaulieu2016FD,Comby2018NatComm,Beaulieu2018NatPhys,Demekhin2018PRL,Goetz2019PRL,Rozen2019PRX,Ordonez2022PCCP} (multi-photon) regime, where the use of two-color light pulses enables unique opportunities for coherent control \cite{Demekhin2018PRL,Goetz2019PRL,Rozen2019PRX,Ordonez2022PCCP}, and has also been recently observed in anions \cite{triptow2023imaging}.

The nonlinear regime of purely electric-dipole light-matter interactions also enables efficient all-optical chiral spectroscopy\cite{Fischer2000PRL,Belkin2000PRL,Belkin2001PRL,Fischer2002CPL,Fischer2003PRL,Fischer2005,Okuno2018JCP,Neufeld2019PRX,Ayuso2019NatPhot,Ayuso2021NatComm,Ayuso2021Optica,Ayuso2022PCCP,Ayuso2022OptExp,Rego2023,Khokhlova2022}, and the same principle can be applied to microwave fields \cite{Patterson2013,Eibenberger2017PRL,Shubert2016,Perez2017,Eibenberger_PRL_2022}, as predicted by Kral, Shapiro and coworkers \cite{Kral_PRL_2001, Kral_PRL_2003} and reformulated for molecular rotations by several groups \cite{lehmann2018influence, leibscher2019principles}.
Chiral sum-frequency generation\cite{Fischer2000PRL,Belkin2000PRL,Belkin2001PRL,Fischer2002CPL,Fischer2003PRL,Fischer2005,Okuno2018JCP} (SFG) is a well established method for chiral recognition.
It requires two incident laser beams with frequencies $\omega_1\neq\omega_2$ propagating in different directions, with wave vectors $\hat{\textbf{k}}_1$ and $\hat{\textbf{k}}_2$, and polarisation $\hat{\textbf{e}}_1$ and $\hat{\textbf{e}}_2$.
The incident fields drive a second-order response in the medium at frequency $\omega_3=\omega_1+\omega_2$ with polarisation $\hat{\textbf{e}}_3=\hat{\textbf{e}}_1\times\hat{\textbf{e}}_2$, which leads to emission of light at this frequency in the direction of $\hat{\textbf{k}}_3=\hat{\textbf{k}}_1+\hat{\textbf{k}}_2$.
SFG is strictly forbidden in the bulk of isotropic media, such as randomly oriented molecules, unless they are chiral.
Since it is driven by purely electric-dipole interactions, the induced signals can be strong if $\omega_3$ is close to resonance \cite{Belkin2001PRL}.
However, the intensity of SFG is not enantiosensitive -- the molecular handedness remains hidden in the phase of the emitted radiation.

To measure the phase of SFG, and thus the medium's handedness, one can make it interfere with a reference signal using a local oscillator \cite{Okuno2018JCP}.
This achiral reference can also be generated from the chiral sample itself, making the near-field intensity enantiosensitive.
To this end, one can take advantage of magnetic interactions\cite{Belkin2000PRL}, or use a constant electric field \cite{Fischer2003PRL}, although these strategies offer limited enantiosensitivity and opportunities for control.

Here we introduce an all-optical approach for efficient chiral discrimination that relies on the interference between two low-order nonlinear processes: chiral sum-frequency generation and achiral third-harmonic generation.
We show how, by sculpting the sub-cycle oscillations of the laser's electric field vector, we can control the ultrafast optical response of the molecules in a highly enantiosensitive manner: quenching the low-order nonlinear response of one molecular enantiomer while maximising it in its mirror twin.
This work shows that  high enantiosensitivity is not limited to highly non-linear processes \cite{Neufeld2019PRX,Ayuso2019NatPhot,Ayuso2021NatComm,Ayuso2021Optica,Ayuso2022PCCP,Ayuso2022OptExp,Rego2023}: it also be achieved in the perturbative regime.

\section*{Results and discussion}

Chiral sum-frequency generation (SFG) can be efficiently driven by any combination of frequencies $\omega_1\neq\omega_2$, as long as $\omega_3=\omega_1+\omega_2$ is close to resonance\cite{Belkin2001PRL}.
Let us impose $\omega_2=2\omega_1$, and thus $\omega_3=3\omega$, with $\omega=\omega_1$ being the fundamental frequency, and consider the next-order nonlinear process: third-harmonic generation (THG).
The medium can efficiently absorb 3 photons of frequency $\omega$ from the first beam, still at relatively low laser intensities, which leads to achiral polarization\footnote{By achiral we mean that the induced polarization has identical amplitude and phase in opposite molecular enantiomers} at frequency $3\omega$.
Can this achiral response of the molecule interfere with the sum-frequency response to produce an enantiosensitive interference?
Unfortunately, momentum conservation dictates that, while the SFG and THG signals have the same frequency, they are emitted in different directions.
Indeed, the THG signal co-propagates with the $\omega$ beam, as $\textbf{k}_{\text{THG}}=3\textbf{k}_{\omega}$, whereas the SFG signal is emitted in between the two driving beams, because $\textbf{k}_{\text{SFG}}=\textbf{k}_{\omega}+\textbf{k}_{2\omega}$.
In the following, we show how, a relatively simple modification of the original SFG setup, allows us to overcome this limitation.

The proposed optical setup combines a linearly polarised beam with frequency $\omega$ and a second beam that carries cross-polarised $\omega$ and $2\omega$ frequencies.
The $\omega$ components are polarised in the plane of propagation, whereas the $2\omega$ component is polarised orthogonal to this plane, see Fig. 1a.
The laser field can be written as:
\begin{align}
\textbf{E} &= \Re\{ \textbf{E}^{(1)} + \textbf{E}^{(2)} \}, \\
\textbf{E}^{(1)} &= E_{\omega}^{(1)} \, e^{-i\omega t + i\textbf{k}_1\cdot\textbf{r}} \, \hat{\textbf{e}}_1, \\
\textbf{E}^{(2)} &= E_{\omega}^{(2)} \, e^{-i\omega t + i\textbf{k}_2\cdot\textbf{r}} \, \hat{\textbf{e}}_2
+ E_{2\omega}^{(2)} \, e^{-2i\omega t -i\phi_{2\omega} + 2i\textbf{k}_2\cdot\textbf{r}} \, \hat{\textbf{e}}_3,
\end{align}
where $E_{\omega}^{(1)}$, $E_{\omega}^{(2)}$ and $E_{2\omega}^{(2)}$ are the electric-field amplitudes including the temporal and spatial Gaussian envelopes, and $\phi_{2\omega}$ is the two-colour phase delay in the second beam.

In our setup (Fig. 1a), we set the ratio between the two input frequencies to $\omega_2/\omega_1=2$, and add the fundamental $\omega$ frequency to the second beam.
These two key features determine a fundamental difference in the properties of the created field with respect to traditional SFG configurations.
In traditional SFG setups, the polarization of the electric-field vector is confined to a plane, and therefore achiral within the electric-dipole approximation.
In our setup, the combination of the two beams creates a locally chiral field: the polarization of the electric-field vector draws a (three-dimensional) chiral trajectory in time \cite{Ayuso2019NatPhot}.

\begin{figure}[h]
\centering
\includegraphics[width=\textwidth]{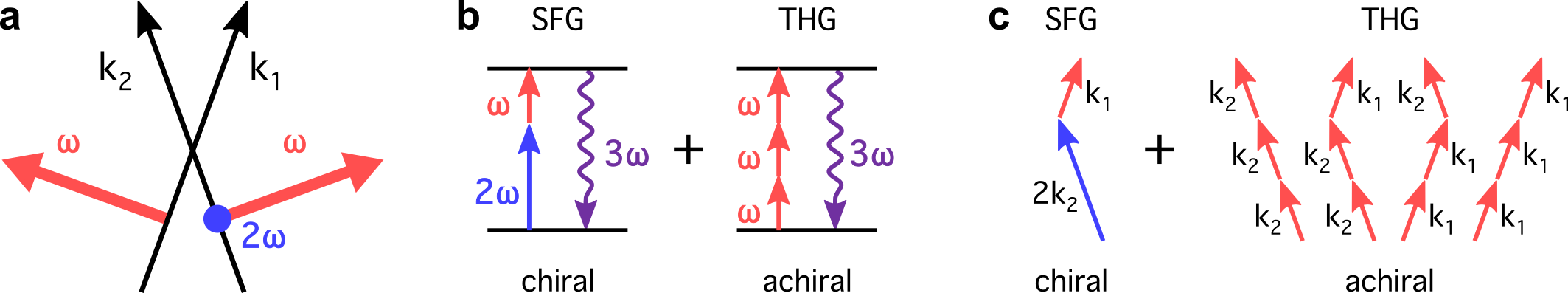}
\caption{\label{fig_setup}
\textbf{enantiosensitive SFG.}
\textbf{a,} The proposed optical setup combines a linearly polarised beam with frequency $\omega$ and a second beam that carries cross-polarised $\omega$ and $2\omega$ frequencies.
The $\omega$ frequency components are polarised in the plane of propagation, the $2\omega$ component is polarised orthogonal to this plane.
\textbf{b,} Multiphoton pathways describing chiral SFG (left) and achiral THG (right). The induced polarisation associated with SFG has the same amplitude and opposite phase in opposite molecular enantiomers, $P_{SFG}^L=-P_{SFG}^R$, whereas the polarisation associated with THG is identical, $P_{THG}^L=P_{THG}^R$.
\textbf{c,} Multiphoton diagrams describing momentum conservation in SFG (left) and THG (right).}
\end{figure}

The generated locally chiral field can drive a strongly enantiosensitive response in a medium of randomly oriented chiral molecules via interference between SFG and THG, see Fig. 1b.
The SFG pathway is as in traditional SFG implementations, see Fig. 1c (left).
However, by adding the $\omega$ frequency to the second beam, we open new THG pathways, see Fig. 1c (right).
The medium can now absorb the three $\omega$ photons from the same beam, or two photons from one beam and one from the other, giving rise to emission of achiral THG in four different directions.
Importantly, one of these pathways, the one involving absorption of one photon from the first beam and two photons from the second beam, leads to achiral THG emission exactly in the same direction as the chiral SFG signal (Fig. 1c).
The two contributions can now interfere, making the intensity of emission strongly enantiosensitive.

To demonstrate our proposal, we have performed state-of-the-art numerical simulations in randomly oriented propylene oxide molecules, see Methods. 
We have considered the following laser parameters:
intensity of the $\omega$ field in the first and second beams $I_{\omega}^{(1)}=I_{\omega}^{(2)}=3 \cdot 10^{12}\,$W/cm$^2$, intensity of the $2\omega$ component $I_{2\omega}^{(2)}=7 \cdot 10^{11}\,$W/cm$^2$, pulse duration $7$ fs, and opening angle $2\alpha=50^\circ$.
Fig. 2a,b shows the intensity of the two contributions to emission at frequency $3\omega$ in the far field (SFG and THG).
As already anticipated, the chiral SFG contribution is emitted at a divergence angle $\arcsin{(\sin(\alpha)/3)}=8.1^\circ$ (Fig. 1c, left), whereas the achiral THG profile shows 4 peaks, at angles $-25.0^\circ$, $-8.1^\circ$, $8.1^\circ$ and $25.0^\circ$, corresponding to the 4 possible $k$-vectors' combinations (Fig. 1c, right).

\begin{figure}[h]
\centering
\includegraphics[width=0.7\textwidth]{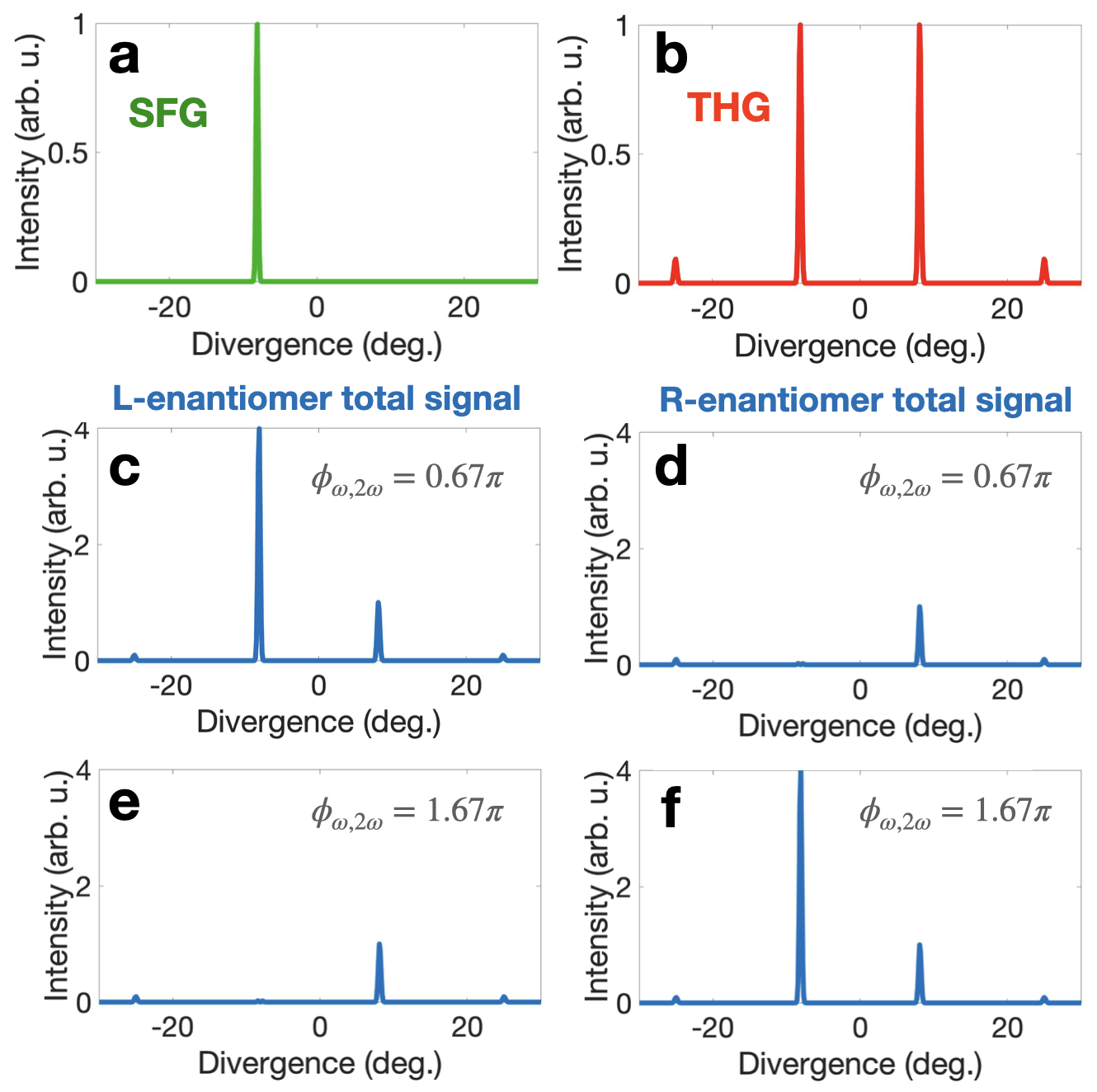}
\caption{\label{fig_far}
\textbf{enantiosensitive SFG in propylene oxide.}
\textbf{a,b,} Intensity at frequency $3\omega$ in the far field associated with chiral SFG (\textbf{a}) and achiral THG (\textbf{b}).
\textbf{c-f} Total intensity emitted from L (\textbf{c,e}) and R (\textbf{d,f}), resulting from adding the chiral SFG (\textbf{a}) and achiral THG (\textbf{b}) contributions, for $\phi_{2\omega}=0.67 \pi $ (\textbf{c,d}) and $\phi_{2\omega}=1.67 \pi$ (\textbf{e,f}).
Laser parameters: $I_{\omega}^{(1)}=I_{\omega}^{(2)}=3 \cdot 10^{12}\,$W/cm$^2$, $I_{2\omega}^{(2)}=7 \cdot 10^{11}\,$W/cm$^2$, pulse duration $7$ fs, and opening angle $2\alpha=50^\circ$.}
\end{figure}

The intensity profiles of the SFG and THG contributions are, individually, not enantiosensitive (Fig. 2a,b).
However, since the SFG contribution is out of phase in opposite molecular enantiomers, the total intensity of emission, resulting from adding the two contributions, becomes strongly enantiosensitive, see Fig. 2c-f.
We have tuned the amplitude of the $2\omega$ component of the driving field so the SFG contribution and the THG contribution at emission angle $-8.1^\circ$ have equal amplitude.
Then, we adjusting the two-colour phase delay in the driving field, we control the phase of SFG, achieving full control over the enantiosensitive interference, and thus over the intensity of emission at frequency $3\omega$.
For $\phi_{2\omega}=0.67\pi$, the SFG and THG contributions interfere constructively in the left-handed molecules (Fig. 2c) and destructively in the right-handed molecules (Fig. 2d).
Changing the two-colour delay by $\pi$, i.e. setting $\phi_{2\omega}=1.67\pi$, changes the phase of the SFG contribution by $\pi$, leading to the opposite effect: suppression from the left-handed molecules (Fig. 2e) and strong emission from the right-handed molecules (Fig. 2f).

Fig. 3 shows how we can fully control the intensity of emission at angle $-8.1^\circ$ in a highly enantiosensitive manner.
Indeed, because the polarisation associated with SFG is out of phase in opposite enantiomers, the values of the two-colour delay that maximise and quench emission from the L (Fig. 3a) and R (Fig. 3b) enantiomers are shifted by $\pi$.
Note that the specific values of the two-colour delay that optimise the enantiosensitive interference depend on the relative phase between the SFG and THG contributions to light-induced polarisation, which record the anisotropy of the chiral molecular potential and thus are molecule-specific quantities.

\begin{figure}[h]
\centering
\includegraphics[width=\textwidth]{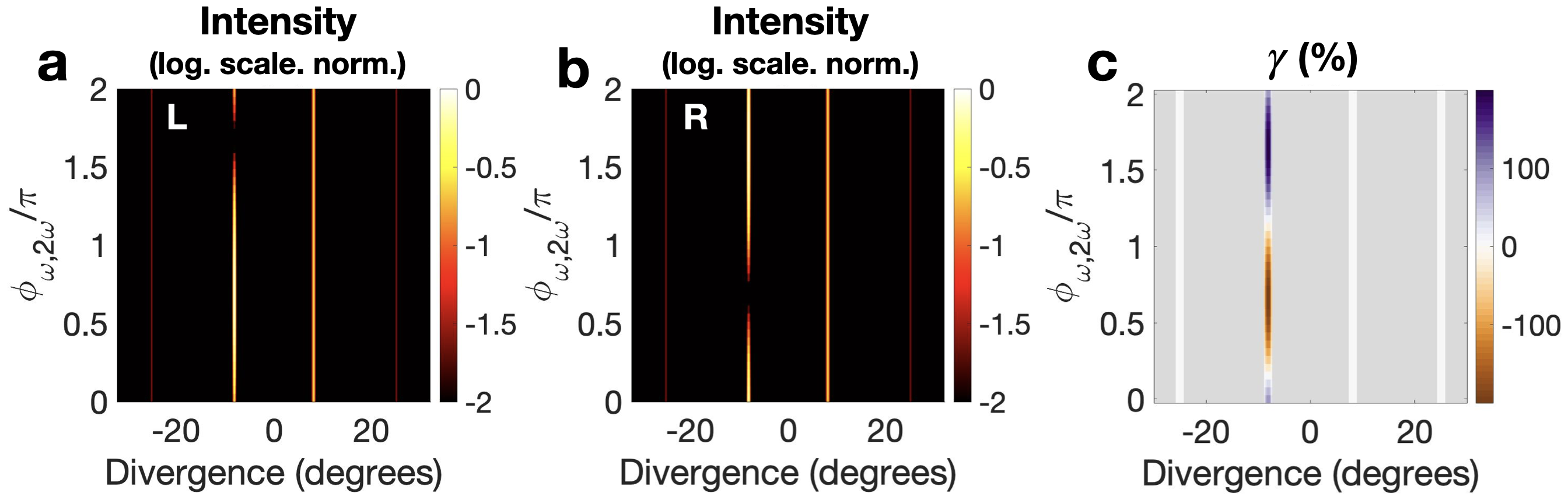}
\caption{\label{fig_control}
\textbf{enantiosensitive control of SFG.}
\textbf{a,b,} Intensity at frequency $3\omega$ emitted from left-handed (\textbf{a}) and right-handed (\textbf{b}) propylene oxide as a function of the divergence angle and the two-colour phase delay.
\textbf{c,} Dissymmetry factor $\gamma=2(I_L-I_R)/(I_L+I_R)$. Grey color indicates no intensity of emission.
See caption of Fig. \label{fig_far} for laser parameters.}
\end{figure}

To quantify the degree of enantiosensitivity in the nonlinear optical response, we use a standard definition of the dissymmetry factor, $\gamma=\frac{I_L-I_R}{(I_L+I_R)/2}$.
As shown in Fig. 3c, $\gamma$ reaches the limits of $\pm200\%$.
That is, we can maximise emission from the left-handed molecules and fully quench it in the right-handed molecules ($\gamma=200\%$), or vice versa ($\gamma=-200\%$), by adjusting the two-colour phase delay in the second beam.
Indeed, because the two-colour delay defines the local handedness of the locally chiral field, it controls the enantiosensitive response of the chiral molecules.

Note that, while the intensity of emission at angle $-8.1^\circ$ is strongly enantiosensitive and can be fully controlled, the peak at $8.1^\circ$ is completely independent on both the molecular handedness and the two-colour delay.
This is due to the different phase matching conditions for SFG and THG, as can be understood in terms of conservation of momentum (Fig. 1), which dictate that the peak at $8.1^\circ$ is solely due to THG.
Thus, this peak constitutes a constant reference which can be used for calibration purposes and for reducing the noise in experimental measures.

\section*{Conclusions}

Our proposal constitutes a simple, yet highly efficient, all-optical approach for chiral discrimination that relies on the interference between low-order nonlinear phenomena (SFG and THG) which are ubiquitous in nonlinear optics.
They have both been widely recorded in chiral and achiral media, in isotropic and anisotropic samples, and in the gas, liquid, and solid phases of matter.
Yet, despite the general importance of chiral molecules, to the best of our knowledge, these two phenomena had not been combined to achieve efficient chiral discrimination.

Our approach takes advantage of the tremendous capabilities of modern optical technology for sculpting the polarization of light with sub-optical-cycle temporal resolution.
Indeed, by controlling the two-colour delay in the proposed optical setup, we can tailor the chirality of the driving field in order to maximise the nonlinear response of a selected molecular enantiomer while suppressing it in its mirror twin.

The possibility of driving strongly enantiosensitive interactions via low-order nonlinear processes creates tremendous opportunities for imaging and controlling molecular chirality on ultrafast timescales using laser fields with gentle intensities, as well as for developing enantiosensitive optical traps and tweezers.

\section*{Acknowledgments}
We acknowledge enlightening discussions with Misha Ivanov, Andr\'es Ordo\~nez, Rose Picciuto, Mary Matthews and Jon Marangos.
L. R. acknowledges financial support from the European Union-NextGenerationEU and the Spanish Ministry of Universities via her Margarita Salas Fellowship through the University of Salamanca;
J. V., L. R. and D. A. acknowledge funding from the Royal Society URF$\backslash$R1$\backslash$201333 and URF$\backslash$ERE$\backslash$210358.
Funded by the European Union (ERC, ULISSES, 101054696). Views and opinions expressed are however those of the author(s) only and do not necessarily reflect those of the European Union or the European Research Council. Neither the European Union nor the granting authority can be held responsible for them.

\section*{Methods}

\subsection*{Single-molecule response of randomly oriented propylene oxide}

The ultrafast electronic response randomly oriented propylene oxide to the proposed driving field was evaluated using real-time time-dependent density functional (TDDFT) theory in Octopus\cite{Marques2003,Castro2006,Andrade2015,Tancogne2020}.
We used the local density approximation\cite{Dirac1930,Bloch1929,Perdew1981} to account for electronic exchange and correlation effects, together with the averaged density self-interaction correction\cite{Legrand2002}.
The 1s orbitals of the carbon and oxygen atoms were described by pseudopotentials.
We expanded the Kohn-Sham orbitals and the electron density onto a spherical basis set of radius $R=41.9$ a.u., with $\Delta x=0.4$ a.u. of spacing between adjacent grid points, and used a complex absorbing potential of width $20$ a.u. and height $-0.2$ a.u. to avoid unphysical reflection effects.
In our TDDFT simulations, we used a two-colour cross-polarised driving field,
\begin{equation}\label{eq_field_simulations}
\mathbf{E}_0(t) = a(t) [ E_{\omega}^{(0)}\cos(\omega t) \hat{\mathbf{x}} + E_{2\omega}^{(0)}\cos(2\omega t) \hat{\mathbf{z}}] ,
\end{equation}
where $E_{\omega}^{(0)}=E_{2\omega}^{(0)}=0.00534$ a.u. are the electric-field amplitudes, $\omega=0.126$ a.u. is the fundamental frequency, and $a(t)$ is sine-squared flat-top envelope of 8 laser cycles of the fundamental frequency, with 2 cycles to rise up, 4 cycles of constant amplitude, and 2 cycles to go down.

We run TDDFT simulations for $208$ different molecular orientations to evaluate the induced polarisation in the randomly oriented ensemble:
\begin{equation}
\mathbf{P}_0 = \frac{1}{8\pi^2} \int_{0}^{2\pi} \int_{0}^{\pi} \int_{0}^{2\pi} \mathbf{P}_{\phi\theta\chi} \; \sin(\theta) \; d\phi \; d\theta \; d\chi,
\end{equation}
where $\phi$, $\theta$ and $\chi$ are the Euler angles, and $\mathbf{P}_{\phi\theta\chi}$ is the polarisation induced in a particular molecular orientation in the laboratory frame.
We used the Lebedev quadrature\cite{Lebedev1999}
of order 7 to integrate over $\phi$ and $\theta$, thus using 26 points to sample the two angles, and the trapezoid method to integrate numerically over $\chi$, using 8 angular points.

\subsection*{Laser field in the interaction region}

The laser field was modelled using Eqs. 1-3 of the main text, with $\mathbf{k}_{1,2}=\frac{2\pi}{\lambda}
[\pm\sin(\alpha)\hat{\mathbf{x}}+\cos(\alpha)\hat{\mathbf{y}}]$, where $\lambda=362$ nm is the fundamental wavelength.
The two beams propagate in the $xy$ plane, at angles $\pm\alpha$ with respect to the $y$ axis.
We assume a thin medium, which could be realised using a flat liquid microjet\cite{Galinis2017,Ferchaud2022}, and thus neglect the spatial modulation of the field properties along the propagation direction, and set the position of the thin sample at $y=0$.
Setting intensity of the $\omega$ frequency component to be the same in the two beams, the total electric-field vector of the laser at $y=z=0$ can be written as
\begin{align}\label{eq_field_x}
\mathbf{E}(x,t) = a(t) [ E_{x}(x) \cos{\left(\omega t\right)} \mathbf{\hat{x}} 
+ E_{y}(x) \sin{\left(\omega t\right)} \mathbf{\hat{y}}
+ E_{z}(x) \cos{\left( 2\omega t + \phi(x) \right)} \mathbf{\hat{z}} ],
\end{align}
with
\begin{align}
E_{x}(x) &= 2 E_{\omega} \cos{(\alpha)}\cos{\left(k\sin{\alpha}\ x \right)} e^{-x^2/w^2}, \\
E_{y}(x) &= 2 E_{\omega} \sin{(\alpha)}\sin{\left(k\sin{\alpha}\ x \right)} e^{-x^2/w^2}, \\
E_{z}(x) &= E_{2\omega} e^{-x^2/w^2}, \\
\phi(x)  &= \phi_{2\omega} + 2k\sin{(\alpha)}\ x,
\end{align}
where $E_{\omega}$ and $E_{2\omega}$ are the electric-field amplitudes, $w$ is the waist of the Gaussian beams, and $\phi_{2\omega}$ is the two-colour phase delay in the second beam.

\subsection*{Nonlinear response in the near field}

The enantiosensitive response of the molecules at frequency $3\omega$ results from the interference between to contributions to light-induced polarisation: chiral sum-frequency generation (SFG) and achiral third-harmonic generation (THG), see Fig. 1 of the main text.
The SFG and THG contributions where obtained by projecting the induced polarisation in the frequency domain $\mathbf{P}_0$ over the $x$ and $y$ axis, 
$\mathbf{P}_0^{\text{THG}}=\mathbf{P}_0(3\omega)\cdot\hat{\mathbf{x}}$ and $\mathbf{P}_0^{\text{SFG}}=\mathbf{P}_0(3\omega)\cdot\hat{\mathbf{y}}$.

Note that this procedure neglects the contribution of the $y$ component of the electric-field vector to both THG and SFG, which reduces significantly the computational cost of the simulations.
As a result, $\mathbf{P}_0^{\text{THG}}$ and $\mathbf{P}_0^{\text{SFG}}$ are linearly polarised along $x$ and $y$, respectively, and not elliptically polarised in the $xy$ plane.
Including the $y$ component of the electric-field vector in the simulations would make both $\mathbf{P}^{\text{THG}}$ and $\mathbf{P}^{\text{SFG}}$ elliptically polarised in the $xy$ plane, but it would not lead to significant changes in the results presented in the main text. 

To model the nonlinear response of the molecules across the interaction region, we assumed that the induced polarisation depends on the field amplitudes according to the number of absorbed photons (see Fig. 1 of the main text), 
\begin{align}
\mathbf{P}^{\text{THG}}(x) &= \mathbf{P}_0^{\text{THG}} \bigg(\frac{E_x(x)}{E_{\omega}^{(0)}}\bigg)^3 , \label{eq_P_THG}\\
\mathbf{P}^{\text{SFG}}(x) &= \mathbf{P}_0^{\text{SFG}} \frac{E_x(x)}{E_{\omega}^{(0)}} \frac{E_z(x)}{E_{2\omega}^{(0)}}
e^{i\phi(x)}, \label{eq_P_SFG}
\end{align}
where $E_{\omega}^{(0)}$ and $E_{2\omega}^{(0)}$ are the electric-field amplitudes used in the TDDFT simulations (Eq. \ref{eq_field_simulations}), and $E_x$, $E_z$ and $\phi$ are the field-amplitudes and phase delay across the transverse coordinate $x$ (Eq. \ref{eq_field_x}).
We performed calculations for the right handed enantiomer of propylene oxide, and the results for the left handed enantiomer were obtained using symmetry arguments.

\subsection*{Enantiosensitive intensity in the far field}

The far-field image was evaluated using Fraunhofer diffraction:
\begin{equation}\label{Fraunhofer}
\mathbf{E}(\kappa) \propto 9\omega^2 \int_{-\infty}^{\infty} \mathbf{P}_{\perp}(x)  \, e^{-iK x} dx,
\end{equation}
where $\kappa$ is the far-field angle (divergence),
$\mathbf{P}_{\perp}$ is the projection of the total polarisation at frequency $3\omega$, $\mathbf{P}=\mathbf{P}^{\text{THG}}+\mathbf{P}^{\text{SFG}}$ (Eqs. \ref{eq_P_THG} and \ref{eq_P_SFG}), onto the plane that is orthogonal to the propagation direction $\hat{\mathbf{k}}=\cos(\kappa)\hat{\mathbf{x}}+\sin(\kappa)\hat{\mathbf{y}}$, 
 and $K=\frac{3\omega}{c}\sin{\left(\kappa\right)}$, with $\omega$ being the fundamental frequency,
 and $c$ being the speed of light in vacuum.

\bibliography{Bibliography}

\end{document}